\documentclass[twocolumn,aps,pra,showpacs,superscriptaddress]{revtex4}
\usepackage{graphicx}
\usepackage{color}
\usepackage{amsmath}
\usepackage{amssymb}

\usepackage[dvips]{hyperref}
\hypersetup{colorlinks=true,linkcolor=blue,citecolor=blue,urlcolor=blue}

\begin{document}

%%%%%%%%%%%%%%%%%%%%%%%%%%%%%%%%%%%%%%%%%%%%%%%%%%%%%%%%%%%%%%%%%%%

\title{Detection efficiency for loophole-free Bell tests with entangled states affected by colored noise}

%Jabaquara Project, paper 2.

%glima@udec.cl,fabio.sciarrino@uniroma1.it,adan@us.es

%%%%%%%%%%%%%%%%%%%%%%%%%%%%%%%%%%%%%%%%%%%%%%%%%%%%%%%%%%%%%%%%%%%

\author{Gustavo Ca\~nas}.
 \affiliation{Departamento de F\'{\i}sica, Universidad de Concepci\'{o}n, 160-C Concepci\'{o}n, Chile}
 \affiliation{Center for Optics and Photonics, Universidad de Concepci\'{o}n, Chile}
 \affiliation{MSI-Nucleus on Advanced Optics, Universidad de Concepci\'{o}n, Chile}

\author{Johanna F. Barra}
 \affiliation{Departamento de F\'{\i}sica, Universidad de Concepci\'{o}n, 160-C Concepci\'{o}n, Chile}
 \affiliation{Center for Optics and Photonics, Universidad de Concepci\'{o}n, Chile}
 \affiliation{MSI-Nucleus on Advanced Optics, Universidad de Concepci\'{o}n, Chile}

\author{Esteban S. G\'omez}
 \affiliation{Departamento de F\'{\i}sica, Universidad de Concepci\'{o}n, 160-C Concepci\'{o}n, Chile}
 \affiliation{Center for Optics and Photonics, Universidad de Concepci\'{o}n, Chile}
 \affiliation{MSI-Nucleus on Advanced Optics, Universidad de Concepci\'{o}n, Chile}

\author{Gustavo Lima}
 \affiliation{Departamento de F\'{\i}sica, Universidad de Concepci\'{o}n, 160-C Concepci\'{o}n, Chile}
 \affiliation{Center for Optics and Photonics, Universidad de Concepci\'{o}n, Chile}
 \affiliation{MSI-Nucleus on Advanced Optics, Universidad de Concepci\'{o}n, Chile}

\author{Fabio Sciarrino}
 %\email{fabio.sciarrino@uniroma1.it}
 %\homepage{http://quantumoptics.phys.uniroma1.it}
 \affiliation{Dipartimento di Fisica, Sapienza
 Universit\`{a} di Roma, I-00185 Roma, Italy}

\author{Ad\'an Cabello}
 %\email{adan@us.es}
 \affiliation{Departamento de F\'{\i}sica Aplicada II, Universidad de
 Sevilla, E-41012 Sevilla, Spain}

%%%%%%%%%%%%%%%%%%%%%%%%%%%%%%%%%%%%%%%%%%%%%%%%%%%%%%%%%%%%%%%%%%%

\date{\today}

%First version: April 2012 (Majadahonda)
%This version: January 11, 2013 (Belo Horizonte), after PRA proofs

%%%%%%%%%%%%%%%%%%%%%%%%%%%%%%%%%%%%%%%%%%%%%%%%%%%%%%%%%%%%%%%%%%%

\begin{abstract}
Loophole-free Bell tests for quantum nonlocality and long-distance secure communication require photodetection efficiencies beyond a threshold $\eta_{\rm crit}$ that depends on the Bell inequality and the noise affecting the entangled state received by the distant parties. Most calculations of $\eta_{\rm crit}$ assume that the noise is random and can be modeled as white noise. However, most sources suffer from colored noise. Indeed, since entangled states are usually created as a superposition of two possible deexcitation paths, a partial distinguishability between the two processes leads to the appearance of colored noise in the generated state. Recently, there was a proposal for a loophole-free Bell test [A.~Cabello and F. Sciarrino, Phys. Rev. X \textbf{2}, 021010 (2012)], where a specific colored noise appears as a consequence of the precertification of the photon's presence through single-photon spontaneous parametric down-conversion. Here we obtain $\eta_{\rm crit}$, the optimal quantum states, and the local settings for a loophole-free Bell test as a function of the amount of colored noise. We consider three bipartite Bell inequalities with $n$ dichotomic settings: Clauser-Horne-Shimony-Holt ($n=2$), $I_{3322}$ ($n=3$), and $A_5$ ($n=4$), both for the case of symmetric efficiencies, corresponding to photon-photon Bell tests, and for the totally asymmetric case, corresponding to atom-photon Bell tests. Remarkably, in all these cases, $\eta_{\rm crit}$ is robust against the colored noise. The present analysis can find application in any test of Bell inequalities in which the dominant noise is of the colored type.
\end{abstract}

%%%%%%%%%%%%%%%%%%%%%%%%%%%%%%%%%%%%%%%%%%%%%%%%%%%%%%%%%%%%%%%%%%%

%03.65.Ud Entanglement and quantum nonlocality
%(e.g. EPR paradox, Bell's inequalities, GHZ states, etc.)
%03.67.Mn Entanglement production, characterization, and manipulation
%42.50.Xa Optical tests of quantum theory

\maketitle

%%%%%%%%%%%%%%%%%%%%%%%%%%%%%%%%%%%%%%%%%%%%%%%%%%%%%%%%%%%%%%%%%%%

\section{Introduction}

%%%%%%%%%%%%%%%%%%%%%%%%%%%%%%%%%%%%%%%%%%%%%%%%%%%%%%%%%%%%%%%%%%%

One of the most surprising predictions of quantum mechanics is that events for which no causal relationship exists (since nothing at the speed of light or slower can connect them) exhibit correlations that do not admit any explanation in terms of local hidden variables \cite{Bell64}. The quest for an incontrovertible experimental confirmation of this prediction is one of the fundamental challenges of modern science. Such confirmation, certified by means of a loophole-free violation of Bell inequalities \cite{Bell64,CHSH69,CH74}, would not only rule out the possibility of describing nature with local hidden variable theories, but would also prove the feasibility of important applications such as secure communication based on physical principles \cite{Ekert91,BHK05,ABGMPS07}.

So far, the results of all the experiments testing Bell inequalities (see, e.g., Refs.~\cite{FC72,ADR82,KMWZSS95,WJSWZ98,RKVSIMW01,MMMOM08,SUKRMHRFLJZ10}) admit explanations in terms of local hidden variables. Assuming that quantum mechanics is correct, the reason why these experiments still do not rule out local hidden variables is simply that none of them satisfies all the conditions under which Bell inequalities are derived. Specifically, they do not simultaneously satisfy the following three conditions: (i) The observers's local measurement choices are independent and random, (ii) one observer's local measurement choice and the other observer's local measurement result are spacelike separated, and (iii) the overall detection efficiency $\eta$, defined as the ratio between detected events and prepared systems, is above a threshold $\eta_{\rm crit}$. Otherwise, the detected events can apparently violate Bell inequalities, whereas the prepared systems do not \cite{Pearle70}. The value of $\eta_{\rm crit}$ depends on the Bell inequality and the state considered. For instance, for the Clauser-Horne-Shimony-Holt (CHSH) Bell inequality \cite{CHSH69,CH74}, $\eta_{\rm crit} \approx 0.67$ for partially entangled states and increases with noise \cite{Eberhard93}.

Hence it is crucial to optimize the strategy, namely, the choice of the Bell inequality and the observables to measure, in order to minimize the required detection efficiency $\eta_{\rm crit}$ for a given experimental platform. This optimization has been previously carried out for white noise \cite{Eberhard93,CL07,BGSS07,VPB10}. However, the large majority of adopted sources suffer from colored noise \cite{CFL05}. Indeed, since entangled state are usually created as a superposition of two possible deexcitation paths, a partial distinguishability between the two processes leads to the appearance of colored noise in the generated state. The partial distinguishability can arise as a consequence of a spectral, temporal, or spatial mismatch between the particle wave packets emitted in the two possible processes. These considerations are true for the generation of photon-photon and hybrid atom-photon entangled states.

Recently, a novel experimental scheme to achieve a loophole-free Bell test adopting a precertification technique was proposed in Ref.~\cite{CS12}. The scheme works as follows. Consider two spatially separated observers, Alice and Bob, and a source between them. The source simultaneously emits two photons $A$ and $B$. Photon $A$ is sent to Alice's location and photon $B$ to Bob's. The key point of the scheme is to precertify the presence of a photon in Alice's (and Bob's) location before she (he) has decided which local measurement will perform. This is achieved by splitting photon $A$ ($B$) into two photons $A1$ and $A2$ ($B1$ and $B2$) by using an enhanced single-photon spontaneous parametric down-conversion. Photon~$A2$ ($B2$) is then detected by a fast nanowire-based superconducting single-photon detector, certifying the presence of photon~$A1$ ($B1$) before the local measurement is fixed. The configuration of the setup is such that photon~$A1$ ($B1$) bears the same information initially encoded in photon~$A$ ($B$). Finally, the local measurement is performed and photon~$A1$ ($B1$) is detected with a superconducting transition-edge sensor (TES), which has a very high detection efficiency. (For details, see Ref.~\cite{CS12}.) The advantage of this scheme with respect to previous proposals is that the Bell test only involves events in which photons~$A2$ and $B2$ are detected. In this way, the transmission losses between the source of photons and Alice and Bob's locations do not affect $\eta$ since the effective transmission efficiency is boosted up to one ($\eta$ is the product of the transmission efficiency and the detector efficiency). Photons~$A1$ and $B1$ can be prepared in a mode highly coupled with the corresponding TESs at the cost of a reduced experiment rate, but without affecting the final detection efficiency. Remarkably, the noise introduced by the precertification process in the final state of photons $A1$ and $B1$ is not random noise that can be modeled as white noise, as it is usually assumed in most calculations of $\eta_{\rm crit}$ \cite{Eberhard93,CL07,BGSS07,VPB10}, but is a colored one.

The goal of this paper is to obtain $\eta_{\rm crit}$, the optimal quantum states, and the local settings for different bipartite Bell inequalities, assuming the specific colored noise characteristic of the precertification scheme. Then we compare these $\eta_{\rm crit}$ values with those for Bell tests affected by white noise and discuss the implications for actual experiments. The paper is organized as follows. In Sec. II we describe the expressions of the quantum states generated while the source is affected by colored noise. The first scenario considered is related to a photon-photon Bell test exploiting the precertification technique. However, it may also correspond to standard Bell tests based on the spontaneous parametric down-conversion source. The second scenario corresponds to an atom-photon Bell test in which the atom is detected with high efficiency and only the photon is affected by colored noise. In Sec.~\ref{Bell inequalities} we describe the three bipartite Bell inequalities considered in our study and explain why we picked them. In Sec.~\ref{Threshold} we explain how $\eta_{\rm crit}$, the optimal states, and the local settings are calculated for the quantum states and Bell inequalities discussed in the preceding sections. The results of these calculations are presented and discussed in Secs.~\ref{Results}--\ref{Results4}.

%%%%%%%%%%%%%%%%%%%%%%%%%%%%%%%%%%%%%%%%%%%%%%%%%%%%%%%%%%%%%%%%%%%

\section{Quantum states in Bell tests with coloured noise}
\label{States}

%%%%%%%%%%%%%%%%%%%%%%%%%%%%%%%%%%%%%%%%%%%%%%%%%%%%%%%%%%%%%%%%%%%

\subsection{Noise for photon-photon Bell tests}
\label{Noise1}

%%%%%%%%%%%%%%%%%%%%%%%%%%%%%%%%%%%%%%%%%%%%%%%%%%%%%%%%%%%%%%%%%%%

We assume that the source initially produces pairs of photons $A$ and $B$ entangled in polarization in the two-qubit entangled state
\begin{equation}
 |\psi\rangle = C |HV\rangle + S |VH\rangle,
\end{equation}
where $|HV\rangle$ is the state in which photon $A$ has horizontal $H$ polarization and photon $B$ has vertical $V$ polarization, $C=\cos(\theta)$, and $S=\sin(\theta)$. As a consequence of a residual distinguishability between the emission of photon pairs with horizontal and vertical polarizations, the initial state of photons $A$ and $B$ is transformed into the following state of photons $A1$ and $B1$:
\begin{equation}
\begin{split}
 \rho=&C^2 |HV\rangle \langle HV| + S^2 |VH \rangle \langle VH| \\
 &+ (1-p)^2 CS \left( |HV\rangle \langle VH|+ |VH\rangle \langle HV|\right),
\end{split}
 \label{ppnoise}
 \end{equation}
where $p$ is the distinguishability between $H$ and $V$. The square factor arises since each photon is affected by a partial distinguishability. Hereafter we will assume that for photon-photon Bell tests, the states \eqref{ppnoise} are the ones reaching the photodetectors.

%%%%%%%%%%%%%%%%%%%%%%%%%%%%%%%%%%%%%%%%%%%%%%%%%%%%%%%%%%%%%%%%%%%

\subsection{Noise for atom-photon Bell tests}
\label{Noise2}

%%%%%%%%%%%%%%%%%%%%%%%%%%%%%%%%%%%%%%%%%%%%%%%%%%%%%%%%%%%%%%%%%%%

In an atom-photon Bell test, the atom is detected with high (ideally perfect) efficiency. Instead of state \eqref{ppnoise}, we consider the resulting state as
\begin{equation}
\begin{split}
 \rho'=&C^2 |HV \rangle \langle HV| + S^2 |VH\rangle \langle VH| \\
 &+ (1-p) CS \left(|HV\rangle \langle VH|+|VH\rangle \langle HV|\right),
\end{split}
 \label{apnoise}
\end{equation}
where we have $(1-p)$, instead of $(1-p)^2$, because only the photon is affected by colored noise. Hereafter we will assume that for atom-photon Bell tests, the states \eqref{apnoise} are the ones reaching the detectors. The difference between states \eqref{ppnoise} and \eqref{apnoise} has been introduced to simplify the connection with the precertification-based loophole-free test introduced in Ref.~\cite{CS12}. Indeed, when an atom-photon scheme is adopted the precertification stage is introduced only in one part of the entangled state, thus reducing the colored noise added to the state.

%%%%%%%%%%%%%%%%%%%%%%%%%%%%%%%%%%%%%%%%%%%%%%%%%%%%%%%%%%%%%%%%%%%

\section{Bell inequalities}
\label{Bell inequalities}

%%%%%%%%%%%%%%%%%%%%%%%%%%%%%%%%%%%%%%%%%%%%%%%%%%%%%%%%%%%%%%%%%%%

Bipartite Bell inequalities are linear combinations of probabilities $P(a,b|x,y)$ of obtaining the result $a$ for the measurement $x$ in Alice's side and the result $b$ for the measurement $y$ in Bob's side, which for any local hidden variables theory have a bound that is violated by the predictions of quantum mechanics. We will focus on three specific tight (i.e., belonging to the minimal set that separates quantum from local correlations \cite{Fine82,Pitowsky89}) bipartite Bell inequalities.

%%%%%%%%%%%%%%%%%%%%%%%%%%%%%%%%%%%%%%%%%%%%%%%%%%%%%%%%%%%%%%%%%%%

\subsection{The CHSH inequality}

%%%%%%%%%%%%%%%%%%%%%%%%%%%%%%%%%%%%%%%%%%%%%%%%%%%%%%%%%%%%%%%%%%%

The CHSH Bell inequality \cite{CHSH69,CH74} has two settings for each party (i.e., $x,y \in \{0,1\}$), each of them with two outcomes (i.e., $a,b \in \{0,1\}$). It is the only tight Bell inequality with two dichotomic settings for each party \cite{Fine82}. It can be written \cite{CH74} as
\begin{equation}
\begin{split}
I_{\rm CHSH}=&P(0,0|0,0)+P(0,0|0,1)+P(0,0|1,0)\\
&-P(0,0|1,1)-P(0,\_|0,\_)-P(\_,0|\_,0)\le0,
\end{split}
\end{equation} where $P(0,\_|0,\_)$ is the marginal probability of obtaining $0$ for the measurement $0$ on Alice's side. In the absence of noise, $\eta_{\rm crit}\approx0.67$ using partially entangled states \cite{Eberhard93} and $\eta_{\rm crit}\approx 0.83$ using maximally entangled states, assuming that all the detectors have the same efficiency \cite{Eberhard93,LS01}. In the case that only Alice's detector has perfect efficiency, $\eta_{\rm crit}=0.50$ using partially entangled states \cite{CL07,BGSS07} and $\eta_{\rm crit}=0.71$ using maximally entangled states \cite{GM87,Larsson98}. The CHSH is the simplest Bell inequality and the bipartite Bell inequality that requires the lowest threshold $\eta$ using qubits to date.

%%%%%%%%%%%%%%%%%%%%%%%%%%%%%%%%%%%%%%%%%%%%%%%%%%%%%%%%%%%%%%%%%%%

\subsection{The $I_{3322}$ inequality}

%%%%%%%%%%%%%%%%%%%%%%%%%%%%%%%%%%%%%%%%%%%%%%%%%%%%%%%%%%%%%%%%%%%

The $I_{3322}$ Bell inequality \cite{Froissart81,Sliwa03,CG04} is the only tight bipartite Bell inequality with three dichotomic settings \cite{CG04}. We will use the asymmetric version of Ref.~\cite{BG08}, which is given by
\begin{equation}
\begin{split}
I_{3322}=&P(0,0|0,0)+P(0,0|0,1)+P(0,0|0,2)\\
&+P(0,0|1,0)+P(0,0|1,1)-P(0,0|1,2)\\
&+P(0,0|2,0)-P(0,0|2,1)\\
&-2P(0,\_|0,\_)-P(0,\_|1,\_)-P(\_,0|\_,0)\le0.
\end{split}
\end{equation}
Our interest in the $I_{3322}$ inequality is due to the fact that, in the absence of noise, $\eta_{\rm crit}=0.43$, assuming that Alice's detectors have perfect efficiency \cite{BGSS07}.

%%%%%%%%%%%%%%%%%%%%%%%%%%%%%%%%%%%%%%%%%%%%%%%%%%%%%%%%%%%%%%%%%%%

\subsection{The $A_5$ inequality}

%%%%%%%%%%%%%%%%%%%%%%%%%%%%%%%%%%%%%%%%%%%%%%%%%%%%%%%%%%%%%%%%%%%

Finally, among the 26 tight bipartite Bell inequalities with four dichotomic settings of Ref.~\cite{BG08}, we will use the one called $A_5$, introduced in Ref.~\cite{IIA06}. Specifically, we will use the following asymmetric version of the $A_5$ inequality:
\begin{equation}
\begin{split}
A_5=&P(0,0|0,1)+P(0,0|0,2)-P(0,0|0,3)\\
&+P(0,0|1,0)+P(0,0|1,1)-P(0,0|1,2)\\
&+P(0,0|1,3)+P(0,0|2,0)+P(0,0|2,2)\\
&+P(0,0|2,3)+P(0,0|3,0)-P(0,0|3,3)\\
&-P(0,\_|0,\_)-P(0,\_|1,\_)-2P(0,\_|2,\_)\\
&-P(\_,0|\_,0)-P(\_,0|\_,1)\le0.
\end{split}
\end{equation}
The $A_5$ inequality is of interest to us because, in the absence of noise and using maximally entangled states, it requires $\eta_{\rm crit}=0.8214$ for the photon-photon scenario, which is (slightly) smaller than required efficiency of the CHSH inequality, the $I_{3322}$, or any of the 26 tight bipartite Bell inequalities with four dichotomic settings of Ref.~\cite{BG08}.

%%%%%%%%%%%%%%%%%%%%%%%%%%%%%%%%%%%%%%%%%%%%%%%%%%%%%%%%%%%%%%%%%%%

\section{Results}

%%%%%%%%%%%%%%%%%%%%%%%%%%%%%%%%%%%%%%%%%%%%%%%%%%%%%%%%%%%%%%%%%%%

\subsection{Method}
\label{Threshold}

%%%%%%%%%%%%%%%%%%%%%%%%%%%%%%%%%%%%%%%%%%%%%%%%%%%%%%%%%%%%%%%%%%%

Once we have the quantum states Alice and Bob will receive (see Sec.~\ref{States}) and the Bell inequalities (see Sec.~\ref{Bell inequalities}), the next step is to modify the Bell inequalities to take into account the nonperfect $\eta$ of the detectors. It is always possible to rewrite Bell inequalities in terms of $\eta$ \cite{Eberhard93,Garuccio95}. The modified Bell inequalities are only violated when $\eta > \eta_{\rm crit}$.

We will assume that the local measurements performed by Alice and Bob are two-outcome von Neumann measurements (i.e., maximal qubit measurements) associated with the following orthogonal states:
\begin{subequations}
 \label{basis1}
 \begin{align}
 |u^{(k)}\rangle_{\phi} &= \cos{\phi} |+^{(k)}\rangle - e^{i \nu_{\phi}} \sin{\phi} |-^{(k)}\rangle,\\
 |v^{(k)}\rangle_{\phi} &= \sin{\phi} |+^{(k)}\rangle + e^{i \nu_{\phi}} \cos{\phi} |-^{(k)}\rangle,
 \end{align}
\end{subequations}
where $k=1$ denotes Alice and $k=2$ denotes Bob. In terms of this basis, the local measurement in which the experimental apparatus has the orientation $\phi$ is represented by the projector $P^{(k)}_{\phi} = |v^{(k)}\rangle_{\phi} \langle v^{(k)}|$.

In the Bell tests based on the CHSH inequality, each party can choose between two different projective measurements (with a total of four measurements, defined in our case by $\phi_1$, $\phi_2$, $\phi_3$, and $\phi_4$). Thus, the CHSH inequality is a two-setting Bell inequality. However, the $I_{3322}$ and $A_5$ are, respectively, three- and four-setting Bell inequalities. The value of $\eta_{\rm crit}$ is a function that depends on the parameters $\{\phi_i\}$ and $\{\nu_{\phi_i}\}$, whose number increases with the number of settings of the Bell inequality. In the case of the CHSH inequality, $\eta_{\rm crit}$ is an eight-variable function, while for the $I_{3322}$ and $A_5$, $\eta_{\rm crit}$ is a 12- and 16-variable function, respectively.

For a given degree of entanglement (i.e., value of $C/S$), level of noise $p$, and Bell inequality, $\eta_{\rm crit}$ has been numerically obtained using the conjugate gradient (CG) method \cite{PTVF92}. The CG method is a heuristic numerical search algorithm that uses the local gradient in a given initial point of the parameter space (defined by $\{\phi_i\}$ and $\{\nu_{\phi_i}\}$) to reach the local minimum point. It converges when the gradient is zero. To map all the local minima for $\eta$ and determine the global minimum (i.e., $\eta_{\rm crit}$), it is necessary to run the CG program for a large and uniform sample of initial points in the parameter space of the corresponding Bell inequality. In order to certify that the global minimum had been actually reached for each of the scenarios, we ran the CG program for samples of more than $10^5$ points for each state considered.

We have used this method to systematically explore the three Bell inequalities for both the photon-photon and the atom-photon scenarios. This provides not only the values of $\eta_{\rm crit}$ for the different scenarios, but also the optimal quantum states and local settings as functions of the degree of entanglement and noise of the states and allows us to compare the cases of colored noise and white noise. We have organized all this information in the following sections.

%%%%%%%%%%%%%%%%%%%%%%%%%%%%%%%%%%%%%%%%%%%%%%%%%%%%%%%%%%%%%%%%%%%

\subsection{Threshold detection efficiencies}
\label{Results}

%%%%%%%%%%%%%%%%%%%%%%%%%%%%%%%%%%%%%%%%%%%%%%%%%%%%%%%%%%%%%%%%%%%
% Fig. 1
%%%%%%%%%%%%%%%%%%%%%%%%%%%%%%%%%%%%%%%%%%%%%%%%%%%%%%%%%%%%%%%%%%%

\begin{figure}[t]
\centerline{\includegraphics[width=0.48\textwidth]{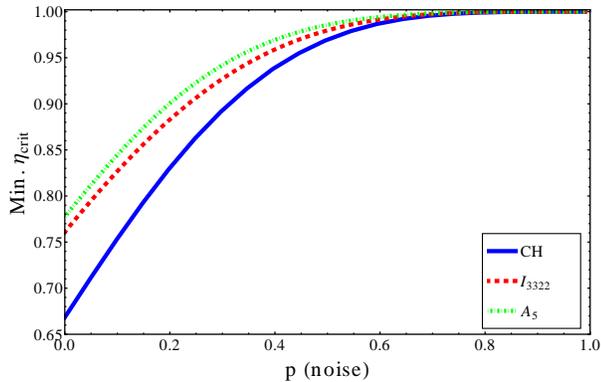}}
\caption{\label{Fig1}Minimum $\eta_{\rm crit}$ for all the states of the form \eqref{ppnoise} with a given level $p$ of colored noise, for the CHSH, $I_{3322}$, and $A_5$ Bell inequalities, under the assumption that all detectors have the same efficiency, as is usually the case in photon-photon Bell tests.}
\end{figure}

%%%%%%%%%%%%%%%%%%%%%%%%%%%%%%%%%%%%%%%%%%%%%%%%%%%%%%%%%%%%%%%%%%%

For the photon-photon Bell test, we have obtained the minimum $\eta_{\rm crit}$ for all the states of the form \eqref{ppnoise} with a given level $p$ of noise, for the three Bell inequalities. The results are shown in Fig.~\ref{Fig1}. They clearly show that the best option for a photon-photon loophole-free Bell test affected by colored noise is the CHSH inequality: Not only does it require the fewest settings, but also the lowest $\eta_{\rm crit}$. For example, for $p < 0.04$, $\eta > 0.70$ is in principle sufficient to certify a loophole-free violation (for more details, see below), while the required $\eta$ is substantially higher using inequalities with more settings.

%%%%%%%%%%%%%%%%%%%%%%%%%%%%%%%%%%%%%%%%%%%%%%%%%%%%%%%%%%%%%%%%%%%
% Fig. 2
%%%%%%%%%%%%%%%%%%%%%%%%%%%%%%%%%%%%%%%%%%%%%%%%%%%%%%%%%%%%%%%%%%%

\begin{figure}[t]
\centerline{\includegraphics[width=0.48\textwidth]{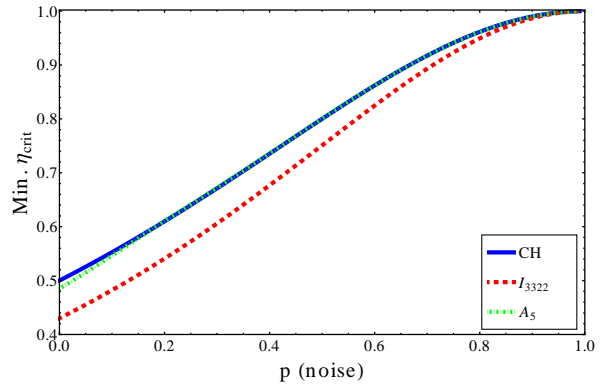}}
 \caption{\label{Fig2}Minimum $\eta_{\rm crit}$ for all the states of the form \eqref{apnoise} with a given level $p$ of colored noise for the CHSH, $I_{3322}$, and $A_5$ Bell inequalities, under the assumption that Alice's detectors are perfect and Bob's have the same efficiency, as is approximately the case in atom-photon Bell tests.}
\end{figure}

%%%%%%%%%%%%%%%%%%%%%%%%%%%%%%%%%%%%%%%%%%%%%%%%%%%%%%%%%%%%%%%%%%%

The conclusion is different for an atom-photon Bell test. Figure~\ref{Fig2} shows the minimum $\eta_{\rm crit}$ for all the states of the form \eqref{apnoise} with a given level $p$ of noise, for the three Bell inequalities. In this scenario, the Bell inequality requiring lower $\eta$ for any $p$ is $I_{3322}$. We have found no benefit in using a Bell inequality with four local settings per party.

%%%%%%%%%%%%%%%%%%%%%%%%%%%%%%%%%%%%%%%%%%%%%%%%%%%%%%%%%%%%%%%%%%%

\subsection{Effect of the colored noise in the threshold efficiencies}
\label{Results2}

%%%%%%%%%%%%%%%%%%%%%%%%%%%%%%%%%%%%%%%%%%%%%%%%%%%%%%%%%%%%%%%%%%%
% Fig. 3
%%%%%%%%%%%%%%%%%%%%%%%%%%%%%%%%%%%%%%%%%%%%%%%%%%%%%%%%%%%%%%%%%%%

\begin{figure}[t]
\centerline{\includegraphics[width=0.48\textwidth]{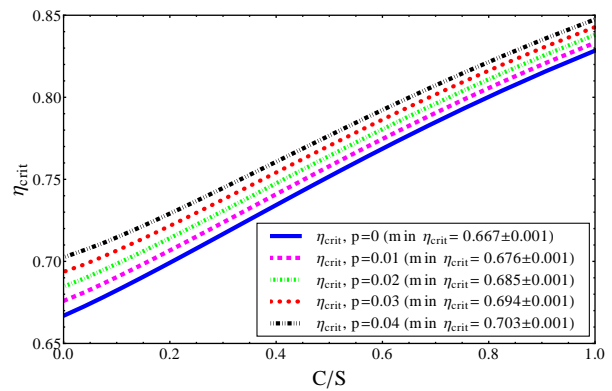}}
 \caption{\label{Fig3}Threshold $\eta_{\rm crit}$ for photon-photon Bell tests of the CHSH inequality using different states of the form \eqref{ppnoise}. The degree of entanglement of the state is denoted by $C/S$: 0 means nonentangled states and 1 maximally entangled states. The minimum $\eta_{\rm crit}$ for each value of $p$ is explicitly shown in the legend.}
\end{figure}

%%%%%%%%%%%%%%%%%%%%%%%%%%%%%%%%%%%%%%%%%%%%%%%%%%%%%%%%%%%%%%%%%%%

The next question is which are the quantum states requiring lower values of $\eta$. Hereafter we will focus on the two most interesting scenarios: the photon-photon test of the CHSH inequality and the atom-photon test of the $I_{3322}$ inequality.

Figure~\ref{Fig3} shows $\eta_{\rm crit}$ for photon-photon Bell tests of the CHSH inequality using different states of the form \eqref{ppnoise}. The entanglement of the states is given by $C/S$: $C=0$ means that the states are product sates and $C/S=1$ means that they are maximally entangled. Figure~\ref{Fig3} shows that weakly entangled states require smaller values of $\eta_{\rm crit}$ than strongly entangled states and adding small levels of colored noise slightly increases $\eta_{\rm crit}$.

%%%%%%%%%%%%%%%%%%%%%%%%%%%%%%%%%%%%%%%%%%%%%%%%%%%%%%%%%%%%%%%%%%%
% Fig. 4
%%%%%%%%%%%%%%%%%%%%%%%%%%%%%%%%%%%%%%%%%%%%%%%%%%%%%%%%%%%%%%%%%%%

\begin{figure}[t]
\centerline{\includegraphics[width=0.48\textwidth]{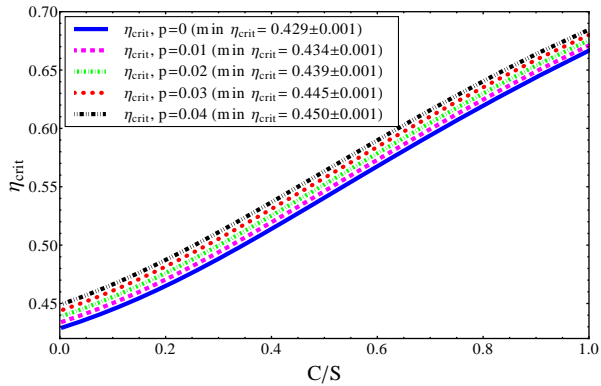}}
 \caption{\label{Fig4}Threshold $\eta_{\rm crit}$ for atom-photon Bell tests of the $I_{3322}$ inequality using different states of the form \eqref{apnoise}. The degree of entanglement of the state is denoted by $C/S$. The minimum $\eta_{\rm crit}$ for each value of $p$ is explicitly shown in the legend.}
\end{figure}

%%%%%%%%%%%%%%%%%%%%%%%%%%%%%%%%%%%%%%%%%%%%%%%%%%%%%%%%%%%%%%%%%%%

Figure~\ref{Fig4} shows that a similar conclusion can be drawn for atom-photon tests of $I_{3322}$ with states of the form \eqref{apnoise}. Again, weakly entangled states require smaller values of $\eta_{\rm crit}$ than strongly entangled ones, and adding small levels of colored noise slightly increases $\eta_{\rm crit}$.

%%%%%%%%%%%%%%%%%%%%%%%%%%%%%%%%%%%%%%%%%%%%%%%%%%%%%%%%%%%%%%%%%%%

\subsection{Optimal settings for a realistic violations}
\label{Results3}

%%%%%%%%%%%%%%%%%%%%%%%%%%%%%%%%%%%%%%%%%%%%%%%%%%%%%%%%%%%%%%%%%%%
% Fig. 5
%%%%%%%%%%%%%%%%%%%%%%%%%%%%%%%%%%%%%%%%%%%%%%%%%%%%%%%%%%%%%%%%%%%

\begin{figure*}[t]
 \centerline{\rotatebox{270}{\includegraphics[width=0.485\textwidth]{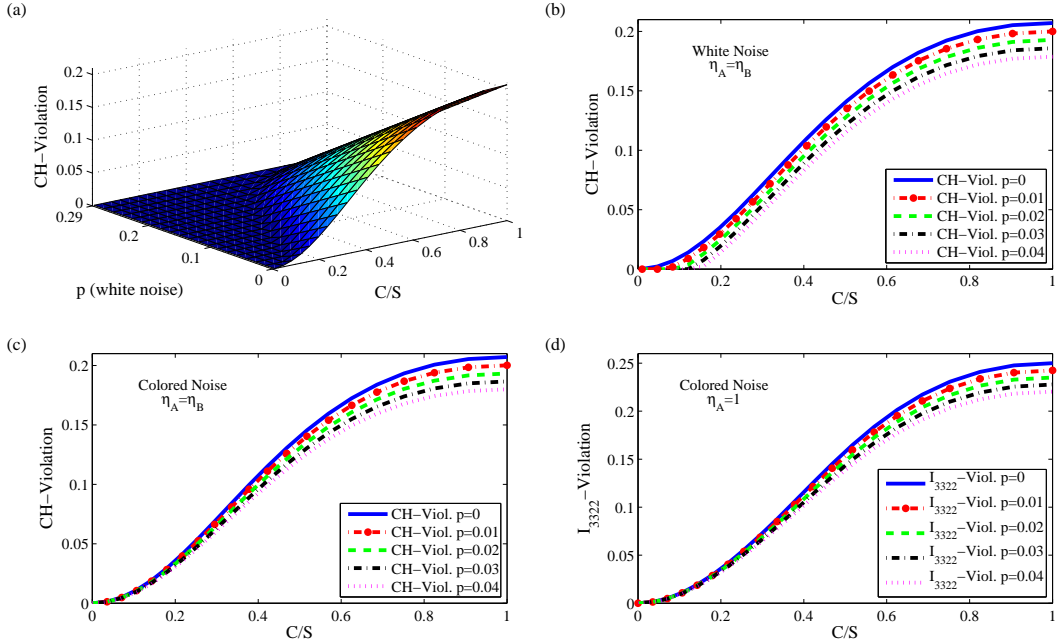}}}
 \caption{\label{Fig5}(a) Value of $I_{\rm CHSH}$ as a function of the white noise $p$ and degree of entanglement $C/S$ of the state for a photon-photon Bell test. (b) Violation of the CHSH inequality for specific values of $p$. (c) Value of $I_{\rm CHSH}$ as a function of the degree of entanglement $C/S$ and for different values of the colored noise $p$ for a photon-photon Bell test. (d) Value of $I_{3322}$ as a function of the degree of entanglement $C/S$ and for different values of the colored noise $p$ for an atom-photon Bell test.}
\end{figure*}

%%%%%%%%%%%%%%%%%%%%%%%%%%%%%%%%%%%%%%%%%%%%%%%%%%%%%%%%%%%%%%%%%%%

Figure~\ref{Fig5} is useful to identify which are the optimal states for a realistic Bell test with colored noise, and also to distinguish the situation from the one in which white noise is assumed. In Fig.~\ref{Fig5}(a) one can clearly see that when the level of white noise increases, then weakly entangled states do not violate the CHSH inequality. Here we assume that a state $|\psi \rangle$ affected by white noise becomes the state $\rho''=(1-p) | \psi \rangle \langle \psi|+\frac {p}{4}\openone$, where $\frac {1}{4}\openone$ denotes the maximally mixed state. Figure~\ref{Fig5}(b) shows the violation of the CHSH inequality for specific values of white noise. Figure~\ref{Fig5}(c) and (d) show that in the case when colored noise is considered, then all entangled states (i.e., all values of $C/S$) violate the inequality. Thus, the optimal states for loophole-free experiments with colored noise are always the very weakly entangled states, since they demand less efficiency and still violate the inequality. This is shown in Fig.~\ref{Fig5}(c) for the photon-photon scenario and the CHSH inequality, and in Fig.~\ref{Fig5}(d) for the atom-photon scenario and the $I_{3322}$ inequality.

Obtaining analytically the local measurements that provide simultaneously maximal violation with minimal $\eta$ as a function of $C/S$ is not a trivial task. In the case of the CHSH inequality, these expressions have been found for the case $p=0$ \cite{LIVLS12}. Here we have obtained numerically optimal states and local settings, assuming a violation that can actually be observed in a Bell test (i.e., a violation greater than $0.01$). Table~\ref{TableI} shows, for two values of $p$, an optimal configuration for a photon-photon Bell test using the CHSH inequality, while Table~\ref{TableII} shows an optimal configuration for an atom-photon Bell test using the $I_{3322}$ inequality. Both tables include the required states (i.e., the value of $C/S$), the expected violation, and the corresponding $\eta_{\rm crit}$. Note that $\eta > 0.73$ guarantees that all the requirements (including a reasonable amount of noise and a noticeable violation) are satisfied for photon-photon Bell test with colored noise. For an atom-photon test $\eta > 0.49$ suffices.

%%%%%%%%%%%%%%%%%%%%%%%%%%%%%%%%%%%%%%%%%%%%%%%%%%%%%%%%%%%%%%%%%%%
% Table I
%%%%%%%%%%%%%%%%%%%%%%%%%%%%%%%%%%%%%%%%%%%%%%%%%%%%%%%%%%%%%%%%%%%

\begin{widetext}
\begin{center}
\begin{table}[h]
\caption{\label{TableI}Local settings $\{\phi_i,\nu_{\phi_i}\}$ for the CHSH inequality in the photon-photon scenario ($\eta_A=\eta_B$).}
\begin{ruledtabular}{
\begin{tabular}{c c c c c c c c c c c c}
%\hline\hline
$p$ & $C/S$ & $I_{\rm CHSH}$ & $\eta_{\rm crit}$ & $\phi_1$ & $\phi_2$ & $\phi_3$ & $\phi_4$ & $\nu_{\phi_1}$ & $\nu_{\phi_2}$ & $\nu_{\phi_3}$ & $\nu_{\phi_4}$ \\\hline
0 & 0.2041 & 0.0362 & 0.6999 & 4.7619 & 1.1651 & 3.0921 & $-0.4057$ & 3.4437 & 3.4437 & 0.3021 & 3.4437\\
0.03 & 0.2041 & 0.0323 & 0.7223 & 1.6180 & 1.1747 & 3.1888 & 0.3961 & 0.9047 & 0.9047 & 0.9047 & 4.0463\\
%\hline \hline
\end{tabular}
} \end{ruledtabular}
\end{table}
\end{center}
%\end{widetext}

%%%%%%%%%%%%%%%%%%%%%%%%%%%%%%%%%%%%%%%%%%%%%%%%%%%%%%%%%%%%%%%%%%%
% Table II
%%%%%%%%%%%%%%%%%%%%%%%%%%%%%%%%%%%%%%%%%%%%%%%%%%%%%%%%%%%%%%%%%%%

%\begin{widetext}
\begin{center}
\begin{table}[h]
\caption{\label{TableII}Local settings $\{\phi_i,\nu_{\phi_i}\}$ for the $I_{3322}$ inequality in the atom-photon scenario ($\eta_A=1$).}
\begin{ruledtabular}{
\begin{tabular}{c c c c c c c c c c c c c c c c}
%\hline\hline
$p$ & $C/S$ & $I_{3322}$ & $\eta_{\rm crit}$ & $\phi_1$ & $\phi_2$ & $\phi_3$ & $\phi_4$ & $\phi_5$ & $\phi_6$ & $\nu_{\phi_1}$ & $\nu_{\phi_2}$ & $\nu_{\phi_3}$ & $\nu_{\phi_4}$ & $\nu_{\phi_5}$ & $\nu_{\phi_6}$ \\\hline
0 & 0.2041 & 0.0462 & 0.4659 & 1.5594 & 4.5618 & 1.2335 & 3.1241 & 6.7563 & 5.6694 & 1.5187 & 4.6603 & 1.5187 & 4.6603 & 4.6603 & 4.6603\\
0.03 & 0.2041 & 0.0433 & 0.4826 & 1.5564 & 4.5733 & 1.2195 & 3.1215 & 6.7253 & 5.6571 & 1.5187 & 4.6603 & 1.5187 & 4.6603 & 4.6603 & 4.6603\\
%\hline\hline
\end{tabular}
} \end{ruledtabular}
\end{table}
\end{center}
\end{widetext}

%%%%%%%%%%%%%%%%%%%%%%%%%%%%%%%%%%%%%%%%%%%%%%%%%%%%%%%%%%%%%%%%%%%

\subsection{Colored noise vs white noise}
\label{Results4}

%%%%%%%%%%%%%%%%%%%%%%%%%%%%%%%%%%%%%%%%%%%%%%%%%%%%%%%%%%%%%%%%%%%
% Fig. 6
%%%%%%%%%%%%%%%%%%%%%%%%%%%%%%%%%%%%%%%%%%%%%%%%%%%%%%%%%%%%%%%%%%%

\begin{figure}[t]
\centerline{\includegraphics[width=0.48\textwidth]{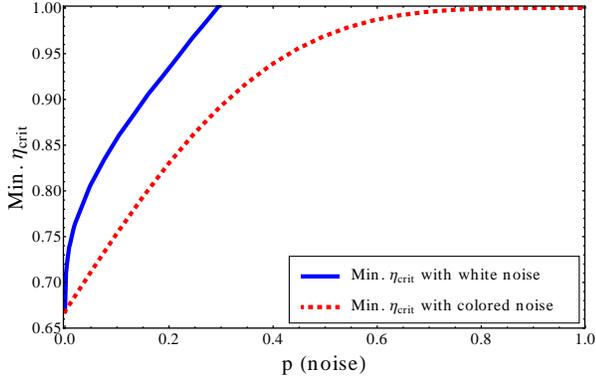}}
 \caption{\label{Fig6}Minimum $\eta_{\rm crit}$ for states with noise $p$ using the CHSH Bell inequality in the photon-photon scenario.}
\end{figure}

%%%%%%%%%%%%%%%%%%%%%%%%%%%%%%%%%%%%%%%%%%%%%%%%%%%%%%%%%%%%%%%%%%%

For a photon-photon Bell test, both for states with white noise and for states with colored noise, the Bell inequality that requires the lowest $\eta_{\rm crit}$ is the CHSH inequality. In Fig.~\ref{Fig6} we compare the value of $\eta_{\rm crit}$ as a function of the level $p$ of colored and white noise for the CHSH inequality. There are two interesting things to observe in this figure. First, the violation of the CHSH inequality is much more robust against colored noise than against white noise. This can be seen by observing that for $p> 0.3$ there is no violation of the CHSH inequality for states with white noise. This is the reason why the curve for $\eta_{\rm crit}$ for states with white noise ends at $p=0.3$. However, in the case of colored noise, there is still a violation of the CHSH inequality, even when $p=1$. A similar observation was made in Ref.~\cite{CFL05}. Second, one can observe that for a given $p$, colored noise demands a lower $\eta_{\rm crit}$ than the one required when white noise is present. This argumentation is valid under the assumption that similar values of colored and white noise can be added to an experiment under a controlled way and independently.

%%%%%%%%%%%%%%%%%%%%%%%%%%%%%%%%%%%%%%%%%%%%%%%%%%%%%%%%%%%%%%%%%%%
% Fig. 7
%%%%%%%%%%%%%%%%%%%%%%%%%%%%%%%%%%%%%%%%%%%%%%%%%%%%%%%%%%%%%%%%%%%

\begin{figure}[t]
\centerline{\includegraphics[width=0.48\textwidth]{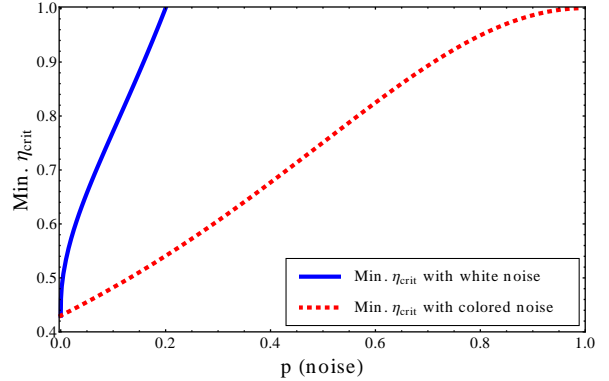}}
 \caption{\label{Fig7}Minimum $\eta_{\rm crit}$ for states with noise $p$ using the $I_{3322}$ Bell inequality in the atom-photon scenario.}
\end{figure}

%%%%%%%%%%%%%%%%%%%%%%%%%%%%%%%%%%%%%%%%%%%%%%%%%%%%%%%%%%%%%%%%%%%

Similar observations can be made for the case of an atom-photon Bell test. There, both for states with white noise and for states with colored noise, the Bell inequality that requires the lowest $\eta_{\rm crit}$ is $I_{3322}$. In Fig.~\ref{Fig7} we compare the value of $\eta_{\rm crit}$ as a function of the level $p$ of colored and white noise for $I_{3322}$. Again, the violation of $I_{3322}$ is much more robust against colored noise than against white noise. Note that the curve for $\eta_{\rm crit}$ for states with white noise ends at $p=0.2$, while for colored noise there is still a violation, even when $p=1$. Moreover, here the difference between $\eta_{\rm crit}$ for a given value of $p$ is even larger than for a photon-photon experiment. For example, for $p=0.2$ a Bell test with entangled states characterized by colored noise requires $\eta_{\rm crit} = 0.55$, while a Bell test with white noise requires $\eta_{\rm crit} = 1$.

%%%%%%%%%%%%%%%%%%%%%%%%%%%%%%%%%%%%%%%%%%%%%%%%%%%%%%%%%%%%%%%%%%%

\begin{figure}[t]
\centerline{\includegraphics[width=0.48\textwidth]{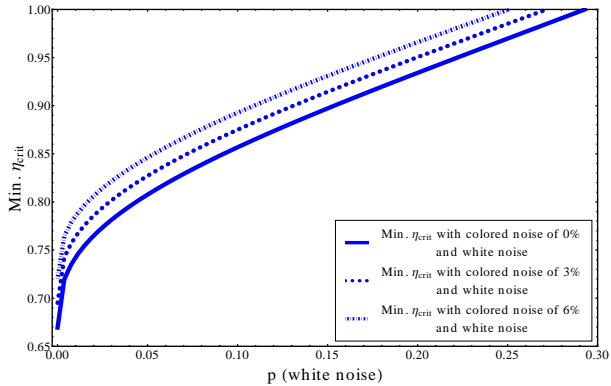}}
 \caption{\label{Fig8}Minimum $\eta_{\rm crit}$ while colored and white noises are present in a CHSH Bell inequality test in the photon-photon scenario.}
\end{figure}

Finally, we consider another realistic scenario where both types of noise are simultaneously present in the experiment. More specifically, we consider the situation where both white and colored noise are present in a photon-photon experiment based on the CHSH inequality. In Fig.~\ref{Fig8} we show the minimum $\eta_{\rm crit}$ for the CHSH Bell inequality, in the photon-photon scenario, for a fixed value of colored noise ($3\%$ and $6\%$) and when white noise is gradually added to the entangled state of Eq.~(\ref{ppnoise}). As expected, one obtains that the robustness of the threshold efficiency degrades fast when white noise is added to the system. Similar results shall be obtained in the case of atom-photon experiments base on the $I_{3322}$ inequality.

%%%%%%%%%%%%%%%%%%%%%%%%%%%%%%%%%%%%%%%%%%%%%%%%%%%%%%%%%%%%%%%%%%%
% Fig. 8
%%%%%%%%%%%%%%%%%%%%%%%%%%%%%%%%%%%%%%%%%%%%%%%%%%%%%%%%%%%%%%%%%%%

%%%%%%%%%%%%%%%%%%%%%%%%%%%%%%%%%%%%%%%%%%%%%%%%%%%%%%%%%%%%%%%%%%%

\section{Conclusions}

%%%%%%%%%%%%%%%%%%%%%%%%%%%%%%%%%%%%%%%%%%%%%%%%%%%%%%%%%%%%%%%%%%%

We have shown that loophole-free Bell tests based on entangled states affected by colored noise exhibit a minimum detection efficiency that is robust for a reasonable amount of noise. This is at variance with the behavior observed for Bell tests in which the noise in the final state is random (white).

For the photon-photon scenario, the most convenient Bell inequality to test is the CHSH inequality using weakly entangled states. We have shown that observable loophole-free violations ($I_{\rm CHSH}=0.0323$) can be achieved with realistic values of the noise ($p=0.03$) and requiring values of $\eta$ that are feasible using the precertification approach ($\eta > 0.7223$). For the atom-photon scenario, the $I_{3322}$ requires an even smaller $\eta$ ($\eta > 0.4826$).

These results support that Bell tests using the photon's precertification may be realistic candidates for a loophole-free Bell test for both the photon-photon and atom-photon scenarios. In addition, the analysis presented here is also valid for other physical systems used to test Bell inequalities when the dominant noise is colored.

%%%%%%%%%%%%%%%%%%%%%%%%%%%%%%%%%%%%%%%%%%%%%%%%%%%%%%%%%%%%%%%%%%%

\begin{acknowledgments}
This work was supported by Projects No.~FIS2008-05596 and No.~FIS2011-29400 (Spain), the Wenner-Gren Foundation (Sweden), FIRB Futuro in Ricerca-HYTEQ (Italy), Project PHORBITECH of the Future and Emerging Technologies program (Grant No.~255914) (EU), FONDECYT 1120067, Milenio P10-030-F and PFB 08024, and CONICYT (Chile).
\end{acknowledgments}

%%%%%%%%%%%%%%%%%%%%%%%%%%%%%%%%%%%%%%%%%%%%%%%%%%%%%%%%%%%%%%%%%%%

%%%%%%%%%%%%%%%%%%%%%%%%%%%%%%%%%%%%%%%%%%%%%%%%%%%%%%%%%%%%%%%%%%%

\end{document}